# cfdmfFTFoam: A front-tracking solver for multiphase flows on general unstructured grids in OpenFOAM

Ehsan Amani[1,*]

[1]Department of Mechanical Engineering, Amirkabir University of Technology (Tehran Polytechnic), Iran

**Abstract**

The Front-Tracking Method (FTM) is a promising approach for numerical solution of multiphase flows, considering a trade-off between accuracy and computational cost. The existing open-source open-access software for FTM is scarce, due to complexity of the coding and algorithms, and is limited to structured Cartesian grids or connectivity-free front mesh hybrid FTMs. To provide a pure FTM solver on general unstructured grids, the Ftc3D FTM code has been integrated into the OpenFOAM CFD software by implementing necessary front mesh to Eulerian grid communication and front nodes advection algorithms, applicable to unstructured grids and for both serial and parallel runs. The new FTM package, called cfdmfFTFoam, has been further equipped with a variety of FTM sub-algorithms, including the front volume correction, remeshing, surface tension computation, indicator function construction, etc. Assessments and validations of the new solver are provided against several standard multiphase flow benchmarks. It is anticipated that cfdmfFTFoam would facilitate future research on and algorithm improvement in the field of FTM.

---

[*] Corresponding author. Address: Mechanical Engineering Dept., Amirkabir University of Technology (Tehran Polytechnic), 424 Hafez Avenue, Tehran, P.O.Box: 15875-4413, Iran. Tel: +98 21 64543404. Email: eamani@aut.ac.ir

# 1. Introduction

Reliable prediction of multiphase flow applications involving complex phenomena, such as Taylor bubble deformation and rise [1], droplet-droplet collision [2], spray breakup [3], boiling [4], bubble nucleation at walls [5], electrowetting [6], alloy solidification [7], contact-line dynamics [8], and Marangoni effect [9], necessitates the use of intricate interface resolving strategies, referred to as Fully Resolved Simulation (FRS) methods, which are divided into two broad categories: the interface tracking, also called Lagrangian or explicit interface advection, and interface capturing, named also as Eulerian or implicit interface advection. Some well-known examples of the first category are the Arbitrary Lagrangian Eulerian (ALE) [9], Immersed Boundary (IB) [10], Front Tracking Method (FTM) [11], Marker And Cell (MAC) [12]; and from the second category, Volume-Of-Fluid (VOF) [11], Level-Set (LS) [13], and Phase-Field (PF) [14] can be cited.

The Open-Access (OA) Open-Source (OS) Computational Fluid Dynamics (CFD) packages for FRS of multiphase flows have been boosting the progress in this area. Probably, the most well-known and commonly-used FRS approach is VOF. Powerful VOF tools are available, including a variety of interface reconstruction algorithms. For instance, the "interFoam" solver of the OpenFOAM (OF) CFD package (www.openfoam.org) offers the algebraic Multidimensional Universal Limiter with Explicit Solution (MULES) [15], and geometric Piecewise Linear Interface Calculation (PLIC) [16] and iso-advector [17] interface reconstruction for VOF on general unstructured grids. The performance of these methods has been extensively compared in the literature, see e.g., [18-21]. The most basic and accurate FRS is ALE. Recently, Schwarzmeier, et al. [22] published an OF-based solver, called "twoPhaseInterTrackFoam", for multiphase flow with soluble and insoluble surfactants. Due to the need for computationally-intensive interface-fitted dynamic mesh algorithms, ALE is limited to canonical problems. To address this issue, IB offers a more computationally efficient solution by removing the necessity for an interface-fitted dynamic mesh, through the solution of fluid flow on an Eulerian grid, while the solid boundaries are tracked in a Lagrangian framework. The effects of the solid boundaries on the fluid are imposed through couplings of forces and velocities. Among OA-OS IB tools, the "sdfibm" solver [23], which is based on the volume-average discrete forcing and applicable to general unstructured grids, can be cited.

FTM can be considered as the extension of IB to multiphase flows where there are fluids flow on the both sides of an interface. In 3D, the interface is tracked by a Lagrangian surface mesh, moving through an Eulerian volumetric grid which is used for the solution of fluid flow equations on the both sides of an interface. The motion and deformation of the interface and interfacial forces on the fluids are accounted for by couplings between the Lagrangian and Eulerian grids. Compared to the interface capturing strategies, like VOF, FTM offers several advantages, including the potential to account for the physics of coalescence and breakup, the straightforward and accurate computation of the interface normal vector and curvature, and avoiding the dissipation error of interface advection. For a recent review on FTM, readers can consult Amani and Tryggvason [24]. There exist several CFD packages with FTM capabilities, including the PARIS code [25], FronTier++ [3, 26] (https://www.ams.sunysb.edu/~chenx/FronTier++_Manual/index.html), TrioCFD [27, 28] (https://github.com/cea-trust-platform/TrioCFD-code), Basilisk [29] (http://basilisk.fr/sandbox/huet/README), all of them are limited to structured Cartesian grids. Recently, a hybrid LS-FTM solver, called lentFoam [30-32] (https://gitlab.com/leia-methods/lent), was publicized for general unstructured grids based on the OF CFD package. Though lentFoam is a powerful tool for multiphase flow FRS, it is not a pure FTM solver and lacks a pure surface Lagrangian mesh representation, and the interfacial force is computed based on the Eulerian field of the LS function. In fact, it is categorized as a connectivity-free FTM approach [33]. Table 1 compares specification of available FTM solvers in more details.

**Table 1** The specification of open-source FTM solvers.

| Software (Type) | Important features | Main References |
|---|---|---|
| PARIS (Pure FTM) | *Eulerian grid:* Structured Cartesian; *Front remeshing/smoothing:* Trapezoidal Sub-grid Undulations Removal (TSUR3D) [34]; *Surface tension computation:* Vertex-based integral formulation [11, 35]; *Indicator construction:* Poisson's Equation (PE) [11, 35] | [11, 25] |
| FronTier (Pure FTM) | *Eulerian grid:* Structured Cartesian; *Surface tension computation:* Least-square quadratic surface fitting [36, 37]; *Topology change:* Improved locally grid-based method [3] | [3, 26] |
| TrioCFD (Pure FTM) | *Eulerian grid:* Structured Cartesian; *Front remeshing/smoothing:* Volume Conserving Smoothing (VCS) [38]; *Indicator construction:* Ray-casting method [39] | [27, 28] |
| Basilisk (Pure FTM) | *Eulerian grid:* Structured Cartesian; *Indicator construction*: PE [11, 35] | [29] |
| lentFoam (Hybrid LS-FTM) | *Eulerian grid:* General unstructured; *Front remeshing/smoothing:* Regularised marching tetrahedra [40]; *Surface tension computation:* LS; *Indicator construction:* LS | [30-32] |
| cfdmfFTFoam (Pure FTM) | *Eulerian grid:* General unstructured; *Front remeshing/smoothing:* TSUR3D [34], VCS [38]; *Surface tension computation:* Element-based direct method [41]; *Indicator construction*: PE [11, 35], Closest Point Transform (CPT) [42]; *Front-to-Field communication*: Reproducing Kernel Particle Method (RKPM) [43-45] | The present work |

To the best of the author's knowledge, no OA-OS CFD tool is available for pure FTM of multiphase flows on general unstructured grids. This work is aimed at filling this gap by introducing and publicizing the "cfdmfFTFoam" solver for 3D general unstructured grids, developed based on the OF CFD tool (www.openfoam.org) and Ftc3D FTM code [25] by the CFDMF group (https://sites.google.com/view/dramani) since 2015. This FTM package can be accessed on CPC Library (…) or the CFDMF repository link (https://github.com/ehsan-amani/cfdmfFTFoam). The main challenges to integrate the Ftc3D FTM code into the unstructured grid solver OpenFOAM include the communication of the flow properties from front Lagrangian surface mesh to Eulerian volumetric fields defined on a general unstructured cell topology in the domain decomposed parallel architecture used in OF. According to Table 1, the Reproducing Kernel Particle Method (RKPM) [43-45] is implemented for this goal. In addition, the original front code, Ftc3D, is enhanced by implementing a variety of new FTM sub-algorithms, including front volume correction, front remeshing/smoothing, surface tension computation, and indicator function construction models (see Table 1).

The rest of the paper is organized as follows: Section 2 presents the mathematical formulation behind FTM, and section 3 discusses the overall structure of the numerical solver. Then, in section 4, the details of the numerical algorithms along with important features of the solver codes are introduced. Section 5 provides several examples of multiphase flow solution procedure and their results to validate the present FTM solver. Finally, the main conclusions are made in section 6.

## 2. Mathematical model

Assuming incompressible flow of $N$ immiscible Newtonian fluid phases, the continuity and momentum, Navier-Stokes (NS), equations can be expressed using the one-fluid formulation [11]:

$$\boldsymbol{\nabla}.\boldsymbol{u} = 0, \tag{1}$$

$$\frac{\partial}{\partial t}(\rho\boldsymbol{u}) + \boldsymbol{\nabla}.(\rho\boldsymbol{u}\boldsymbol{u}) = -\boldsymbol{\nabla}p + \boldsymbol{\nabla}.\{\mu[\boldsymbol{\nabla}\boldsymbol{u} + (\boldsymbol{\nabla}\boldsymbol{u})^T]\} + \rho\boldsymbol{g} + \boldsymbol{F}_\sigma, \tag{2}$$

where $\rho$, $\boldsymbol{u}$, $p$, $\mu$, $\boldsymbol{g}$, and $\boldsymbol{F}_\sigma$ indicate the (one-fluid) density, velocity vector, pressure, dynamic viscosity, gravitational acceleration vector, and surface tension (per unit volume), respectively. The linear rule is usually used to close the one-fluid properties, $\rho$ and $\mu$,

$$\rho = \sum_{q=1}^{N} I_q \rho_q, \qquad (3)$$

$$\mu = \sum_{q=1}^{N} I_q \mu_q, \qquad (4)$$

where $I_q(\mathbf{x}, t)$ is the $q^{\text{th}}$-phase indicator function governed by:

$$\frac{\partial I_q}{\partial t} + \mathbf{u} \cdot \nabla I_q = 0; \; q = 1, \ldots, N-1, \qquad (5)$$

$$I_N = 1 - \sum_{q=1}^{N-1} I_q. \qquad (6)$$

The surface tension per unit (fluid) volume, $\mathbf{F}_\sigma$, in Eq. (2) is expressed in terms of the interfacial force per unit (interface) area, $\mathbf{f}_\sigma$,

$$\mathbf{F}_\sigma = \int_{A_f} \delta(\mathbf{x} - \mathbf{x}_f) \mathbf{f}_\sigma dA_f, \qquad (7)$$

where $\delta(\mathbf{x})$ indicates the 3D delta function, $\mathbf{x}_f$ the front location, and the integration is over the interface (front) area, $A_f$. The location of each point at the front, $\mathbf{x}_f$, is tracked using the following Lagrangian equation:

$$\frac{d\mathbf{x}_f}{dt} = \mathbf{u}_f. \qquad (8)$$

Assuming no interfacial mass transfer, $\mathbf{u}_f = \mathbf{u}(\mathbf{x} = \mathbf{x}_f)$. The solution of the pure advection equations, Eqs. (5), while preserving the sharpness of the interface, is a serious challenge in CFD. Here, FTM is used to address this problem and is discussed in detail in section 3.

## 3. Numerical method: Overview

In a 3D FTM solver, Eqs. (1)-(4) are solved on an Eulerian volumetric grid, while the solution of Eqs. (5) is avoided on this grid due to its sensitivity to the inherent dissipation error of the Eulerian discretization. Instead, a Lagrangian surface mesh, called front, is used to track the interface between phases. Then, the indicator function and interfacial forces are computed given the front location. Figure 1 shows a sample of front and Eulerian grid.

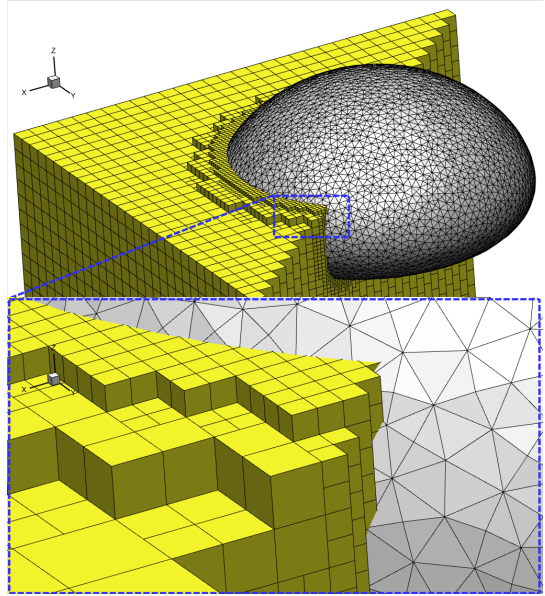

**Figure 1** An FTM Lagrangian surface mesh (front) with triangular cells shown in gray within a cut of Eulerian volumetric grid with hexahedral cells colored in yellow (top). A zoomed view of the two levels of non-interface-fitted dynamic Eulerian mesh refinement near the front (bottom).

Figure 2 represents the flowchart of the present FTM solver, cfdmfFTFoam. The solver has been developed in OF Foundation CFD package (www.openfoam.org) version 9. The governing equations are coupled by a PISO-SIMPLE (Pressure-Implicit with Splitting of Operators - Semi-Implicit Method for Pressure-Linked Equations) or PIMPLE algorithm [46]. For the front tracking at each time step, six consecutive steps are executed: front volume correction, front remeshing, surface tension computation (at the front), front-to-field communication, indicator function construction, and front advection. The details of these algorithms, implemented in cfdmfFTFoam, are given in sections 4.1-4.6.

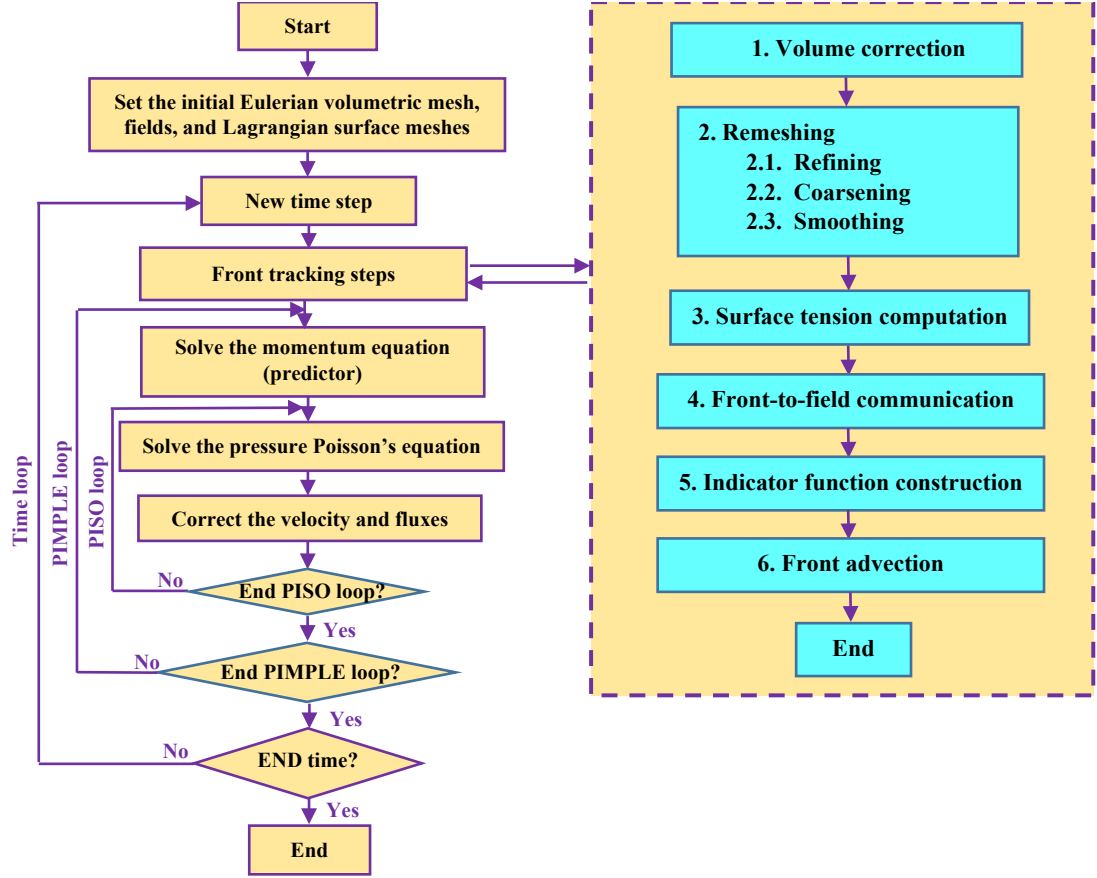

**Figure 2** The cfdmfFTFoam solver flowchart: The steps of the PIMPLE pressure-velocity coupling (left) and the FTM sub-algorithms at each time step (right).

## 4. The code structure and numerical algorithm details

The present FTM package is composed of a solver, "cfdmfFTFoam", a library, "frontTracking", and a series of tutorial cases, "tutorials", including the ones used for the solver validation in section 5. The package file structure, class hierarchies, code description, and Unified Modeling Language (UML) diagrams are detailed in Appendix A. Next, we describe the key elements of the present FTM, outlined in Figure 2.

### 4.1. Volume correction

FTM, like some other FRSs such as LS, is not inherently mass-conserving (or volume-conserving in case of incompressible flows) and needs *ad hoc* algorithms to preserve this property. A volume correction algorithm [44, 47, 48], called VC2 here, is available in the package. Based on VC2, to preserve the volume, the position of each front vertex or point, $x_p$, is corrected by

$$x_p^c = x_p + \epsilon n_p. \tag{9}$$

Here, the subscript $p$ refers to a front point (or vertex), and $n_p$ is the front unit normal vector at the vertex location, approximated by

$$n_p = \frac{A_p}{\|A_p\|}, \tag{10}$$

$$A_p = \sum_{\substack{m=1 \\ p \in m}}^{N_{E,p}} [(x_p - x_{p+}) \times (x_p - x_{p-})], \tag{11}$$

where $\|.\|$ returns the magnitude of a vector, and the summation is over all triangular elements $m$ connected to vertex $p$, for which, the other two vertices of element $m$ are chosen in the counterclockwise order, i.e., $(p-, p, p+)$. The model constant, $\epsilon$, in Eq. (9) is determined by the solution of the following cubic equation:

$$(V_0 - V) = a\epsilon^3 + b\epsilon^2 + c\epsilon, \tag{12}$$

where $V$ and $V_0$ are the current and initial front-enclosed volumes, respectively, which are computed by

$$V = \sum_{m=1}^{N_E} V_m, \tag{13}$$

$$V_m = \frac{1}{3}[x_{c,m} \cdot S_m], x_{c,m} = \frac{x_I + x_{II} + x_{III}}{3}, \tag{14}$$

$$S_m = A_m n_m = \frac{1}{2}(x_{II} - x_I) \times (x_{III} - x_I), \tag{15}$$

where for each front triangular element $m$, $A_m$ is the element area, $n_m$ the unit normal vector, and the element vertices $I$, $II$, and $III$ are shown in Figure 15b. The summation in Eq. (13) is over all elements of a front. The coefficients of Eq. (12) are computed by

$$a = \frac{1}{6} \sum_{m=1}^{N_E} [n_I \cdot (n_{II} \times n_{III})],$$

$$b = \frac{1}{6} \sum_{m=1}^{N_E} [x_I \cdot (n_{II} \times n_{III}) + x_{II} \cdot (n_{III} \times n_I) + x_{III} \cdot (n_I \times n_{II})], \tag{16}$$

$$c = \frac{1}{6} \sum_{m=1}^{N_E} [n_I \cdot (x_{II} \times x_{III}) + n_{II} \cdot (x_{III} \times x_I) + n_{III} \cdot (x_I \times x_{II})].$$

The steps of VC2 are shown in Algorithm 1. To give an example of code implementation of FTM sub-algorithms, the details of the code corresponding to Algorithm 1 is explained in Appendix B.

---

**Algorithm 1: VC2**

---

Run for all **bubbleData**s

1: Compute $V$ by Eqs. (13)-(15)
2: **if** $|V - V_0|/V_0 > VETolerance$ **then**     ▷ $VETolerance = 0.001$ (default)
3:     **for each** $p \in vertexList$ **do**
4:         $\mathbf{n_p} \leftarrow$ Compute $\mathbf{n_p}$ for each vertex by Eqs. (10) and (11)
5:     **end for**
6:     **for each** $m \in elementList$ **do**
7:         **Collect:** the contribution of the terms in the brackets in Eq. (16)     ▷ Collect cubic equation constants
            to the coefficients $a$, $b$, and $c$
8:     **end for**
9:     $a = a/6; b = b/6; c = c/6;$
10:    Solve Eq. (12) for $\epsilon$
11:    **for each** $p \in vertexList$ **do**
12:        $\mathbf{x}_p^c = \mathbf{x}_p + \epsilon \mathbf{n}_p$     ▷ vertex position correction
13:    **end for**
14: **end if**

---

### 4.2. Remeshing

Remeshing is required to control the number and quality of front elements. In the present FTM package, this is done by three consecutive sub-algorithms: refining, coarsening, and smoothing (or undulation removal) (see Figure 2 and Figure 15). The refining and coarsening algorithms available in the FTM library have been converted to C++ from Ftc3D Fortran code developed by Tryggvason and coworkers [11, 25, 35, 49]. The refining step is sketched in Figure 3a, and its algorithm is presented in Algorithm 2. In line 2 of Algorithm 2, the refining criteria for a triangular element $E$ are checked, where $l_{\max} = \max(l_1, l_2, l_3)$, $l_{\min} = \min(l_1, l_2, l_3)$ (see Figure 15b), $AR$ is the element aspect ratio computed by

$$AR = \frac{\sqrt{3}}{4A}\left(\frac{l_1 + l_2 + l_3}{3}\right)^2, \quad (17)$$

and $A$ is the element surface area. $\Delta_{\max}$, $\Delta_{\min}$, $AR_{\max}$ are the prespecified element maximum edge length, minimum edge length, and maximum allowable aspect ratio thresholds. The location of the new vertex, $P_{\text{NEW}}$, is computed by

$$\mathbf{x}_{\text{NEW}} = 0.5(\mathbf{x}_1 + \mathbf{x}_2) + 0.125(\mathbf{x}_3 + \mathbf{x}_4) - 0.0625(\mathbf{x}_5 + \mathbf{x}_6 + \mathbf{x}_7 + \mathbf{x}_8). \quad (18)$$

The coarsening step is given in Algorithm 3 and Figure 3b.

### Algorithm 2: Refining

Run for all `bubbleData`s

1: **for** each $m \in elementList$ **do**
2:    **if** $l_{max} > \Delta_{max} || (AR > AR_{max} \ \&\& \ l_{min} < \Delta_{min})$ **then**     ▷Refining criteria
3:       $E_I \leftarrow$ find the neighboring element sharing $l_{max}$
4:       Label the surrounding vertices by $P_1$ to $P_8$ according to Figure 3a (left)
5:       Add a new vertex, $P_{NEW}$, with a position computed by Eq. (18)
6:       $P_1$ and $P_2$ lose their connection to each other and connect to $P_{NEW}$ instead
7:       $P_{NEW}$ is added to the connections of $P_3$ and $P_4$
8:       $P_1$ is replaced with $P_{NEW}$ in the vertices of the two original elements (Figure 3a (middle))
9:       Two additional elements, $E_{NEW1}$ and $E_{NEW2}$, are added to the element list (Figure 3a (right))
10:    **end if**
11: **end for**

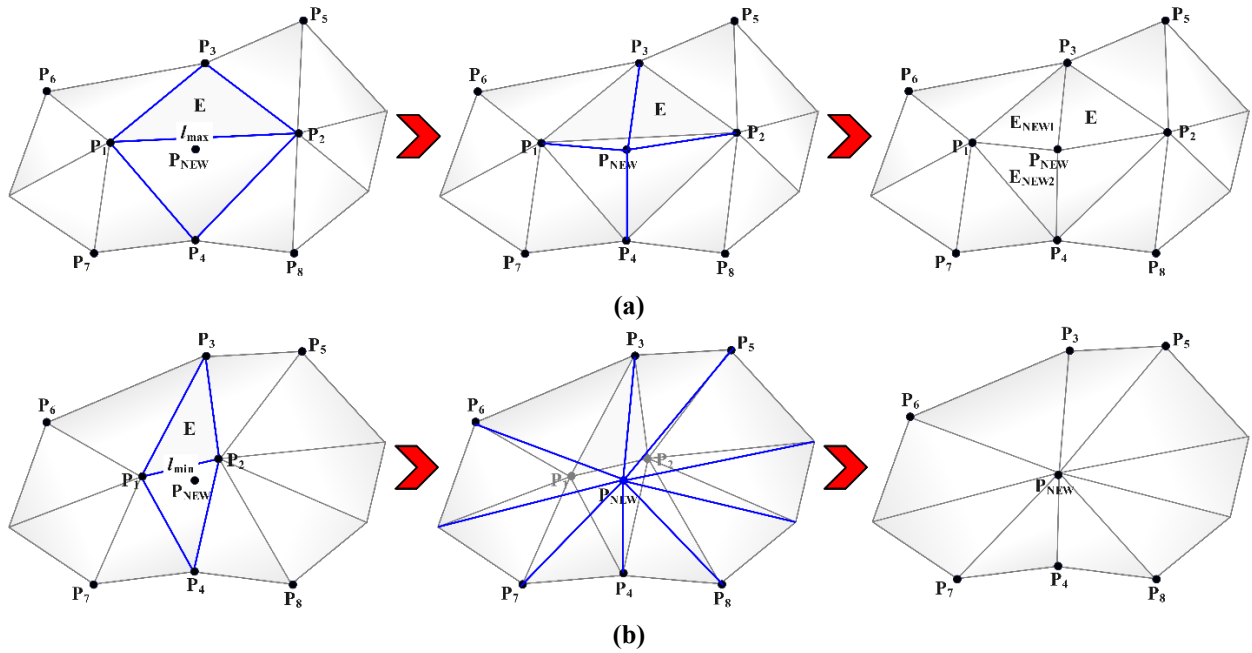

**Figure 3** (a) Refining process, and (b) Coarsening process.

### Algorithm 3: Coarsening

Run for all `bubbleData`s

1: **for** each $m \in elementList$ **do**
2:    **if** $l_{min} < \Delta_{min}$ **then**     ▷Coarsening criterion
3:       $E_I \leftarrow$ find the neighboring element sharing $l_{min}$
4:       Label the surrounding vertices by $P_1$ to $P_8$ according to Figure 3b (left)
5:       Add a new vertex, $P_{NEW}$, with a position computed by Eq. (18)
6:       The position of $P_1$ is updated as $x_1 = x_{NEW}$
7:       All vertices connected to $P_2$ update their connections to $P_1$ (Figure 3b (middle))
8:       $P_2$ is replaced with $P_1$ in the vertices of all elements using $P_2$ (Figure 3b (middle))
9:       Two original elements are removed from the element list (Figure 3b (right))
10:    **end if**
11: **end for**

The smoothing or undulation removal step aims to remove surface mesh noises and undulations by the movement of front vertices while keeping the mesh connectivity unchanged. For the smoothing step, three different algorithms have been implemented in the present FTM library, including TSUR3D [34], and two 3D variants of VCS algorithms [38], called VCSIII and VCSIV. The details of these algorithm are available in the original references or in the appendices of the recent review by Amani and Tryggvason [24].

*4.3. Surface Tension computation*

For calculation of the surface tension imposed on each front element, i.e., the discretized counterpart of $f_\sigma dA_f$ in Eq. (7), the direct element-based method (`directElementBased`) proposed by Deen, et al. [41] has been implemented, where the net surface tension on element $m$ are computed by

$$f_{\sigma,m} A_m = \frac{1}{2}\sigma \sum_{e=1}^{3}(t_{me} \times n_e) \tag{19}$$

Here, $A_m$ is the surface area of the central triangular element $m$ in Figure 4 (calculated by Eq. (15), $A_m = \|S_m\| = \|A_m n_m\|$). The other vectors in Eq. (19) are shown in Figure 4 and can be simply computed given element unit normal vectors, $n_m$, and vertices positions.

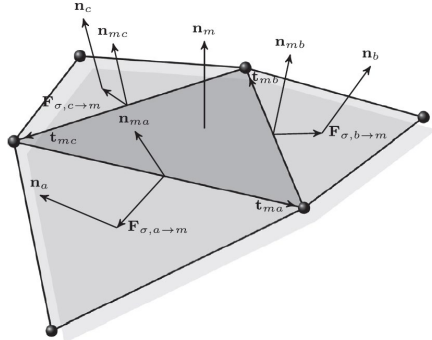

**Figure 4** The direct Element-Based method [50]. In this figure, elements $a$, $b$, and $c$ correspond to elements 1, 2, and 3 in our notation.

*4.4. Front-to-field communication*

To transfer data, e.g., the computed surface tension at each front element, from a front to an Eulerian volume mesh, RKPM [43-45, 51] is adopted. In RKPM, the discretized form of Eq. (7) is used:

$$F_\sigma(x_j) = \sum_{k=1}^{N_E} f_{\sigma,k} A_k \, \tilde{g}_k\left(\frac{x_j - x_k}{h_k}\right), \tag{20}$$

where $x_j$ and $x_k$ indicate the center of the j[th] Eulerian grid node and k[th] front element, respectively. $h_k$ is the characteristic length scale at the k[th] front element, and the summation is over all front elements. Three methods are available in the package to determine $h_k$: `fixed`, which uses a given fixed value for this parameter, `bubbleAverage`, which uses the average value of length scales of all Eulerian cells ($\Delta = V_j^{1/3}$, $V_j$ is the j[th] Eulerian grid cell volume) within a cuboid box enclosing the front, and `frontAverage`, using the average length scale of Eulerian grid cells in the spheres of influence of all front nodes. The RKPM kernel function for the k[th] front element, $\tilde{g}_k(x)$, is expressed by

$$\tilde{g}_k(x) = (\beta_0 + \beta_1 x + \beta_2 y + \beta_3 z)\tilde{\delta}_k(x), \tag{21}$$
$$\tilde{\delta}_k(x) = \tilde{\delta}_k(x)\tilde{\delta}_k(y)\tilde{\delta}_k(z), \tag{22}$$

where $\tilde{\delta}_k(x)$ is an approximation to the 1D delta function, chosen here as the Peskin's function [11, 52]:

$$\tilde{\delta}_k(x) = \begin{cases} \frac{1}{2\alpha h_k}\left[1 + \cos\left(\frac{\pi|x|}{\alpha}\right)\right] & ; |x| < \alpha \\ 0 & ; \text{otherwise} \end{cases}, \tag{23}$$

The parameter $\alpha h_k$ determines the radius of the kernel sphere of influence, where $\alpha$ (or `hFactor` in case files) is by default $\alpha = 2$. For each front element ($k$), the coefficients $\beta_0$ to $\beta_3$ in Eq. (21) are determined by solving a system of equations:

$$\begin{pmatrix} m_{0,0,0} & m_{1,0,0} & m_{0,1,0} & m_{0,0,1} \\ m_{1,0,0} & m_{2,0,0} & m_{1,1,0} & m_{1,0,1} \\ m_{0,1,0} & m_{1,1,0} & m_{0,2,0} & m_{0,1,1} \\ m_{0,0,1} & m_{1,0,1} & m_{0,1,1} & m_{0,0,2} \end{pmatrix} \begin{pmatrix} \beta_0 \\ \beta_1 \\ \beta_2 \\ \beta_3 \end{pmatrix} = \begin{pmatrix} 1 \\ 0 \\ 0 \\ 0 \end{pmatrix}, \tag{24}$$

where the moment $m_{a,b,c}$ is calculated by

$$m_{a,b,c} = \sum_{j=1}^{N_{j,k}} \left(\frac{x_j - x_k}{h_k}\right)^a \left(\frac{y_j - y_k}{h_k}\right)^b \left(\frac{z_j - z_k}{h_k}\right)^c \tilde{\delta}\left(\frac{x_j - x_k}{h_k}\right) V_j. \tag{25}$$

Here, the sum is over all $N_{j,k}$ Eulerian grid cells in the sphere of influence of the k[th] front element. The details of the implemented RKPM algorithm are reported in Algorithm 4.

`RKPM` preserves the zeroth- and first-order moments (forces and torques) during force distribution. A simpler, more commonly-used algorithm in FTMs, which can be selected by the `basic` keyword in the case setups, conserves the force only and can be simplified by setting $\beta_1 =$ $\beta_2 = \beta_3 = 0$ in the RKPM equations. It is worth mentioning that `RKPM` is equivalent to `basic` for uniform Cartesian grids.

**Algorithm 4: RKPM**

Input: $f_{\sigma,k} A_k$ (computed in section 4.3)
Initialization: set $F_\sigma(x_j)$ to zero, then run for all `bubbleData`s

1: **for each** $m \in elementList$ **do**
2:     $x_m \leftarrow$ Center of element $m$
3:     $N_{j,k} \leftarrow$ Find cells within the sphere of influence of $x_m$
4:     **for each** $j \in N_{j,k}$ **do**
5:         $x_j \leftarrow$ Center of cell $j$
6:         **Collect:** $m^p_{a,b,c}$ by Eq. (25)
7:     **end for**
8:     $\beta_i \leftarrow$ Solve the linear system Eq. (24)
9:     **for each** $j \in N_{j,k}$ **do**
10:        $x_j \leftarrow$ Center of cell $j$
11:        Compute $\tilde{g}_k[(x_j - x_k)/h_k]$ by Eqs. (21)-(23)
12:        **Collect:** $F_\sigma(x_j)$ by Eq. (20)
13:    **end for**
14: **end for**

It should be noted that the coefficient matrix in Eq. (24) could become singular in extreme conditions, e.g., when the kernel sphere of influence contains few Eulerian grid nodes, depending on the grid configuration and/or the value of the coefficient $\alpha$. To handle this situation, the Tikhonov regularization is incorporated by adding a small value ($10^{-15}$) to the diagonal components of the matrix.

*4.5. Indicator function construction*

Two algorithms are available for the indicator function, $I_q$, construction in the present FTM package: `poisson` and `cpt`. In the former, a Poisson's Equation (PE) is solved [11, 35, 49]:

$$\nabla^2 I_q = \boldsymbol{\nabla} \cdot \boldsymbol{G}_I, \tag{26}$$

where $\boldsymbol{G}_I$ is the indicator function gradient on the Eulerian grid. For computing $\boldsymbol{G}_I$, it is sufficient to replace $\boldsymbol{F}_\sigma$ with $\boldsymbol{G}_I$ and $f_{\sigma,k} A_k$ with $n_k A_k$ in Algorithm 4. The latter method, `cpt`, stands for Closest Point Transform (CPT) [42], and is based on the distance, $\varphi = \varphi(x_i)$, of each node of the Eulerian grid, $x_i$, to its nearest front element. The indicator function of phase $q$ is computed by

$$I_q(\pmb{x}_i) = \begin{cases} 1; & \varphi > \gamma, \\ \frac{1}{2}\left[1 + \frac{\varphi}{\gamma} + \frac{1}{\pi}\sin\left(\frac{\pi\varphi}{\gamma}\right)\right]; & -\gamma \leq \varphi \leq \gamma, \\ 0; & \varphi < -\gamma, \end{cases} \quad (27)$$

where $\gamma = \alpha h_k$, and $\alpha$ and $h_k$ can be defined like those in the RKPM method (section 4.4). It is worth nothing that the PE method generally provides a smoother indicator function variation over the interface and is superior to geometric methods like CPT, especially on general unstructured grids. The latter method needs implementing several post-filtering operations to remove the noise in the indicator function field and avoid solution instability.

*4.6. Front Advection*

Since the "frontTracking" library has been built on the OF "lagrangian" library, all functionalities required for the front advection, including the particle tracking, boundary treatment, fluid velocity interpolation at a front vertex location ($\pmb{u}_f$), and ODE integration to solve Eq. (8) are inherited from the standard OF. OF is equipped with an intricate face-to-face procedure [53] for Lagrangian particle tacking in general unstructured grids and domain-decomposition parallel processing.

## 5. Results

This section is aimed at validating the cfdmfFTFoam solver results and providing several examples of case setup with this solver. In the rest of this text, the following settings are used, unless stated otherwise: Volume correction (`vc2`), Remeshing ($\Delta_{\min}/\Delta$= 0.25, $\Delta_{\max}/\Delta$= 0.6, $AR_{\max} = 1.5$, `vcsiv` with a frequency of 50 time steps), surface tension computation (`directElementBased`), front-to-field communication (`RKPM`), indicator construction (`poisson`), front advection (trilinear interpolation+1st-order explicit Euler). The case files including the details of the numerical settings are available in the "tutorials" folder of the package.

*5.1. 3D deformation benchmark*

To test the FTM volume correction, remeshing, and front advection algorithms, avoiding the uncertainties of the other parts (see Figure 2), including solving the NS equations, surface tension computation, front-to-field communication, and indicator construction, we consider the test of front deformation in a 3D divergence-free analytical velocity field proposed by Leveque [54] as

$$u(x,y,z,t) = 2\cos\left(\frac{\pi t}{T}\right)\sin^2(\pi x)\sin(2\pi y)\sin(2\pi z),$$
$$v(x,y,z,t) = -\cos\left(\frac{\pi t}{T}\right)\sin(2\pi x)\sin^2(\pi y)\sin(2\pi z), \tag{28}$$
$$w(x,y,z,t) = -\cos\left(\frac{\pi t}{T}\right)\sin(2\pi x)\sin(2\pi y)\sin^2(\pi z).$$

A spherical droplet of diameter $d$ is placed in this velocity field at position $x_0$ and time $t = 0$. For this test, the reference solution by Gorges, et al. [55] is chosen where $d = 0.3$, $x_0 = (0.35, 0.35, 0.35)$, $T = 3$, $d/\Delta = 13.5$, and $\Delta t = 2.5 \times 10^{-4}$. For this simulation, the well-known tsur3d undulation removal algorithm with the frequency of 200 is adopted. The choice of this frequency depends on the selected smoothing algorithm and the physics of the specific case under consideration. As a rule of thumb, this parameter value is opted for to be large enough to suppress small-scale numerical undulations on the front (of the order of the front cell size) while maintaining the steepest physical curvature of the interface. This may need a trial-and-error approach and a sensitivity analysis. A simple sample of such analyses is reported for the fourth benchmark in section 5.4. For a more systematic and detailed analysis, the readers can consult our previous publication [48]. The computational domain is chosen to be a cube with unit edge length, which is discretized by uniform cubic cells of size $\Delta$. A comparison of the drop deformation predicted by the present FTM solver and the reference FTM solution is illustrated in Figure 5a,b. For additional assessments, we performed VOF simulations with the algebraic MULES interface reconstruction [15] using the standard "interFoam" solver of OF9 (Figure 5c) and geometric isoAdvector interface reconstruction [17] using the standard "interIsoFoam" solver of OF-v1912 (Figure 5d). The numerical parameters of the interface reconstruction are chosen similar to the cases provided by Gamet, et al. [20]. Figure 5 highlights an advantage of FTM over implicit interface advection methods, like VOF and LS. Using FTMs (Figure 5a,b), in contrast to VOFs (Figure 5c,d), the unwanted and uncontrolled interface rupture and breakup are avoided when the drop undergoes large stretching and interface segments approach each other to distances smaller than the local Eulerian grid size. As a quantitative measure, the advection error can be computed by

$$\varepsilon_A = \frac{100}{V_d}\sum_{j=1}^{N_j}|I_d(x_j,T) - I_d(x_j,0)|V_j \;(\%), \tag{29}$$

where $V_d$ is the drop volume, and the summation is over all Eulerian grid cells. This error quantifies the difference between the drop interface locations at time instances $t = 0$ and $t = T$. The values of this error are $\varepsilon_A = 1\%$, 41%, and 36% for the present FTM, VOF-MULES, and VOF-isoAdvector solutions, respectively.

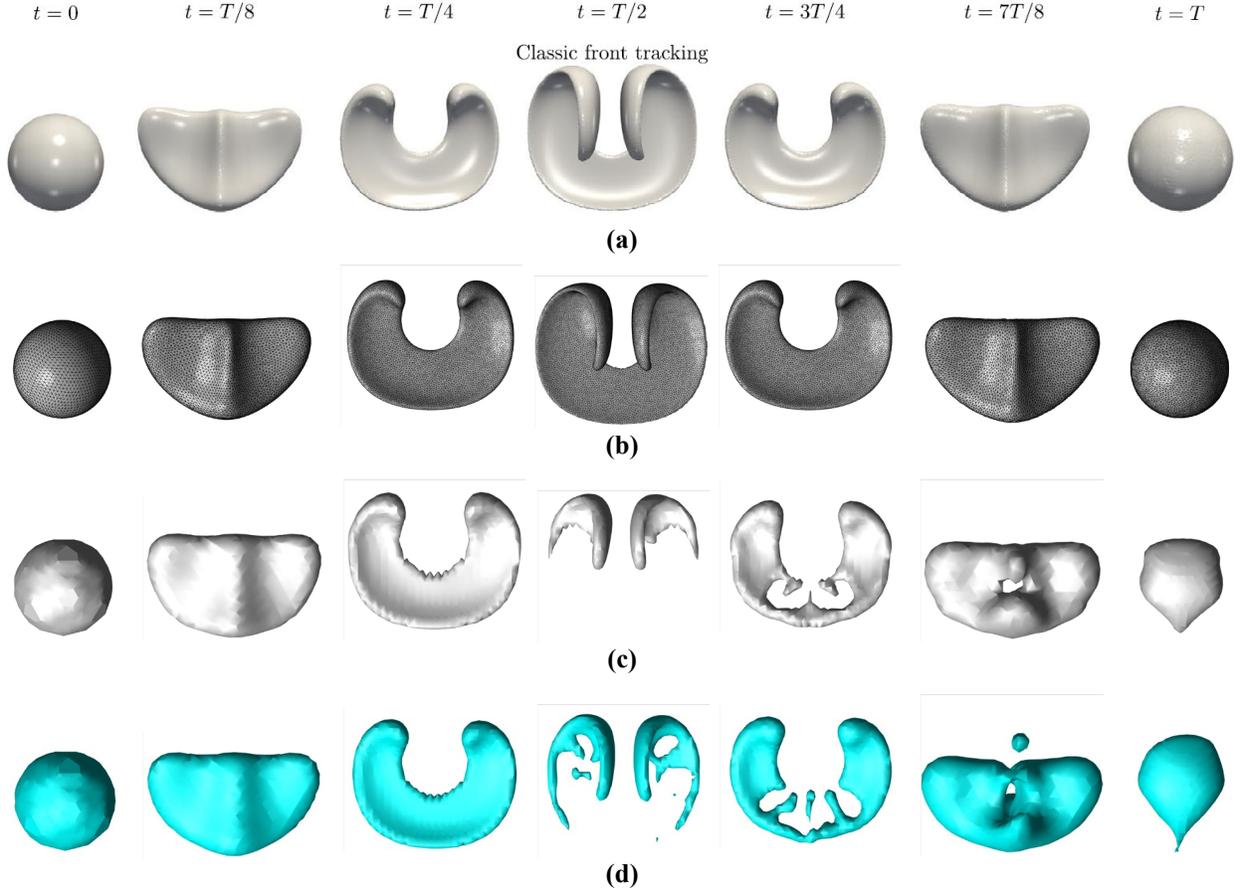

**Figure 5** The 3D deformation field benchmark: Droplet deformation in one period $t = 0$ to $t = T = 3$. The standard FTM by Gorges, et al. [55] (a) and the present solutions ($d/\Delta = 13.5$) using "cfdmfFTFoam" (b), "interFoam-MULES" (c), and "interIsoFoam-isoAdvector" (d).

To report a grid study and evaluate numerical convergence of the present FTM solver, simulations with four different grid resolutions are performed, and an important integral parameter, namely the drop sphericity at $t = T/2$, is computed for these simulations. The drop sphericity is defined by $\emptyset(t) = \pi^{1/3}\big(6V_d(t)\big)^{2/3} A_d^{-1}(t)$, where $V_d(t)$ and $A_d(t)$ are the drop volume and surface area, respectively. The sphericity measures the departure from a spherical interface ($\emptyset = 1$). The apparent order of convergence, $p$, Richardson extrapolated error, $e_{\text{ext}}$, and Grid Convergence Index (GCI), computed based on the formulations given in reference [56], are reported in Table 2. According to this data, the mean apparent order of convergence (of the two sets of grids) is about 2, and the mean uncertainty of the drop shape sphericity computation is below 1%.

To assess the performance of the volume correction algorithm VC2, described in section 4.1, a test is conducted without the volume correction algorithm (none). The volume conservation error,

$$\varepsilon_V = \frac{100}{V_d(0)}[V_d(t) - V_d(0)] \ (\%), \tag{30}$$

with and without the volume correction algorithm versus time are reported in Figure 6. The volume correction algorithm effectively resolves the inherent deficiency of the explicit interface method FTM in conserving mass or volume. While, for this test case, the absence of a volume correction algorithm results in a small error (below 0.2%), in some applications with strong near wall effects and/or large velocity gradients, it can lead to much larger errors, see, e.g., [48].

**Table 2** The GCI computation for two sets of triplet grids.

| Parameter | The 1st triplet (Grids 1, 2, 3) Values | The 2nd triplet (Grids 2, 3, 4) Values |
|---|---|---|
| $(d/\Delta)_1, (d/\Delta)_2, (d/\Delta)_3$ | 21.6, 17.1, 13.5 | 17.1, 13.5, 10.8 |
| $N_1, N_2, N_3$ | 373248, 185193, 91125 | 185193, 91125, 46656 |
| $r_{32}, r_{21}$ | 1.3, 1.3 | 1.3, 1.25 |
| $\emptyset_1, \emptyset_2, \emptyset_3$ | 0.243682, 0.244256, 0.245265 | 0.244256, 0.245265, 0.246749 |
| $p, e_{ext}^{21}$ | 2.35, 0.32% | 1.93, 0.72% |
| $GCI^{21}$ | 0.40% | 0.89% |

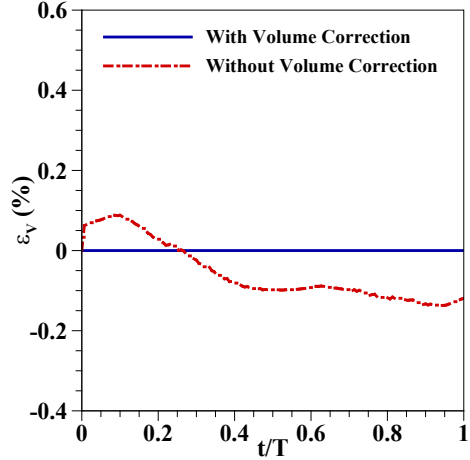

**Figure 6** The 3D deformation field benchmark: The effect of using volume correction algorithm (VC2). The volume conservation error versus the dimensionless time.

The present FTM front remeshing algorithm controls the size and the quality of the triangulation (the aspect ratio). To show the performance of this algorithm, a quality measure of the front mesh is reported over time in Figure 7. As can be observed from the coloring of the triangles in this figure, the predefined threshold of $AR \leq 1.5$ is met at all times. It is worth nothing that the present remeshing parameters ($\Delta_{min}/\Delta = 0.25, \Delta_{max}/\Delta = 0.6, AR_{max} = 1.5$) are chosen based on the recommendations provided by Tryggvason and coworkers, e.g., [11, 25, 48].

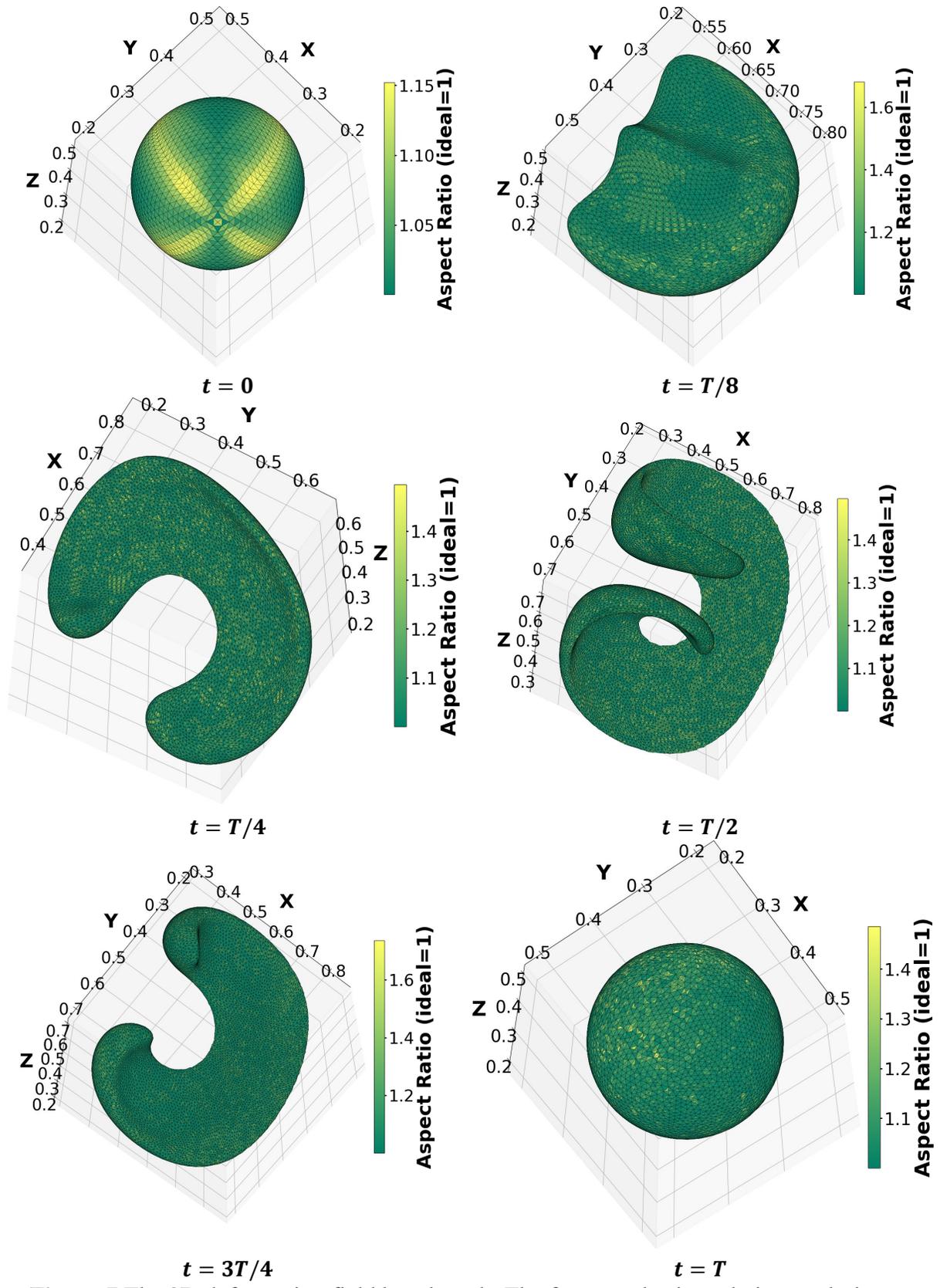

**Figure 7** The 3D deformation field benchmark: The front mesh triangulation evolution over time with the contours of the cell aspect ratio.

*5.2. Stagnant droplet benchmark*

To assess the surface tension computation and front-to-field communication sub-algorithms, the well-known 3D stagnant droplet benchmark is used. Here, a spherical droplet of diameter $d = 0.8$ is placed at the center of a box with an edge length of $L = 2$. A case with a Laplace number of $\text{La} = \rho_d d\sigma/\mu_d^2 = 12000$, densities of $\rho_d = \rho_c = 1$, viscosities of $\mu_c = 1$ and $\mu_d = 8.165 \times 10^{-3}$, surface tension coefficient of $\sigma = 1$, and numerical settings of $d/\Delta = 25.6$ and $\Delta t = 10^{-4}$ is selected similar to the main setup by Gamet, et al. [20]. Note that since the bubble should be a closed surface in cfdmfFTFoam, we simulated the whole 3D box with slightly different boundary conditions, compared to the 1/8th symmetry used in the referenced work. The simulation continues to a stationary state, and the results are reported in Figure 8. The solution accuracy can be assessed by the deviation of the calculated drop-interior and -exterior pressure difference, $\Delta p = p_{\text{int}} - p_{\text{ext}}$, from the analytical solution and the strength of the parasitic currents, which are measured, respectively, by

$$\varepsilon_p = 100 \left|\frac{d\Delta p}{4\sigma} - 1\right| \text{ (\%)}, \tag{31}$$

$$\text{Ca}_{\text{sc}} = \frac{\mu_d u_{\max}}{\sigma}, \tag{32}$$

where $u_{\max}$ is the maximum of the velocity field magnitude. The result of the "cfdmfFTFoam" FTM solver for this benchmark is reported in Figure 8a. The error measures for this solution are $\varepsilon_p = 0.046$ (%) and $\text{Ca}_{\text{sc}} = 2.1 \times 10^{-3}$. For further assessment, we performed a simulation with the "interFoam" solver using the algebraic MULES, with the same computational grid, boundary conditions, and similar numerical settings as those of the FTM. The error measures for this solution are $\varepsilon_p = 12.6$ (%) and $\text{Ca}_{\text{sc}} = 3.1 \times 10^{-3}$. A comparison of the evolution of the parasitic current over time using different solvers is provided in Figure 8b. These results advocate the improvement by the present FTM solver over conventional VOF approaches.

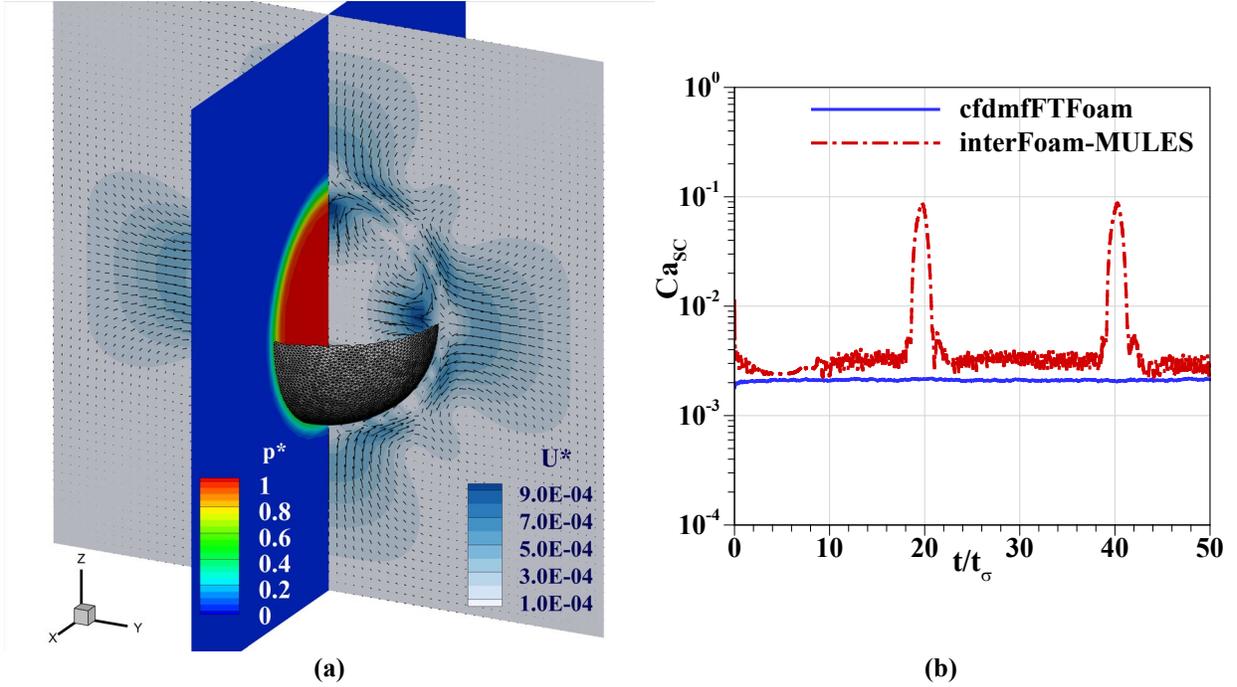

(a)          (b)

**Figure 8** The 3D stagnant drop benchmark: a) The dimensionless pressure ($p^* = d(p - p_{\text{ext}})/(4\sigma)$) contours on the $y = 0$ plane and parasitic current vectors and dimensionless velocity magnitude ($U^* = \mu_d U/\sigma$) contours on the $x = 0$ plane. The lower-half of the front is shown. b) The spurious current Capillary number versus the dimensionless time ($t/t_\sigma$ where $t_\sigma = \sqrt{\rho d^3/\sigma}$). A comparison of the "cfdmfFTFoam" and "interFoam-MULES" solvers.

### 5.3. Droplet in Poiseuille flow benchmark – Stokes regime

The next benchmark is deformation of a droplet in a Poiseuille flow within a circular tube at low Reynolds numbers (the Stokes regime), where an analytical solution is available for the terminal droplet shape [57]. An initially spherical droplet of diameter $d$ drifts along the axis of a circular tube of diameter $D$ and length $L$ under a Poiseuille flow condition with a bulk velocity of $u_b$. For this case, the governing parameters are chosen as Re = $\rho_c u_b D/\mu_c = 0.1$, Ca = $\mu_c u_b/\sigma = 1$, $d/D = 0.3$, $\rho_c/\rho_d = 1$, $\mu_c/\mu_d = 1$, and $L/D = 6$. To show the capability of the solver for unstructured Eulerian grids, meshes with different cell topologies are chosen, including hexahedral (Figure 9b), hexagonal-prism (Figure 9c), polyhedral (Figure 9d), and tetrahedral (Figure 9e) cells, all with $d/\Delta_e = 20$, where $\Delta_e$ is the average grid-cell edge size (for a cubic cell, $\Delta_e = \Delta$).

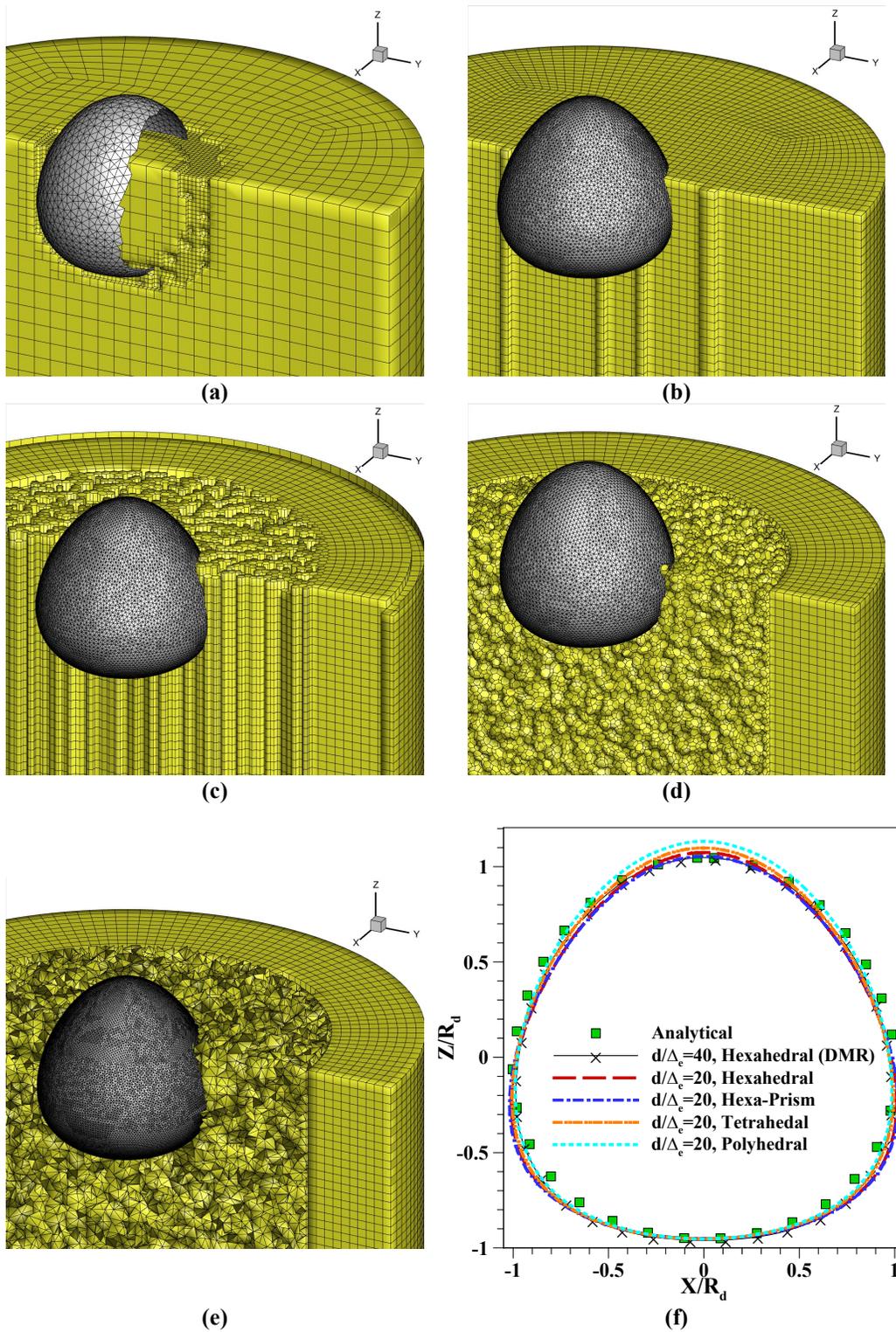

**Figure 9** The droplet in a Poiseuille Stokes flow benchmark: A cut of the front and the Eulerian grid for a) hexahedral-DMR, b) hexahedral, c) hexahedral-prism, d) polyhedral, and e) tetrahedral cell topologies. f) A comparison of the terminal drop shapes predicted by the present "cfdmfFTFoam" solver and analytical solution [57].

A finer resolution of $d/\Delta_e = 40$ is achieved using another hexahedral grid with 2 levels of Dynamic Mesh Refinement (DMR) near the drop interface (Figure 9a). To reduce the total number of grid cells, all grids have a block-structured mesh region near the walls far from the droplet. Figure 9f shows that there is reasonable agreement between the analytical solution for the terminal drop shape and the FTM predictions using all grid topologies.

To show parallel scalability of the new solver, simulations with different number of processors are conducted and the parallel efficiency and speedup are reported in Figure 10. For a reference, these metrics are also shown for the original interFoam solver of OF9. OF uses a domain decomposition architecture. Based on this data, the efficiency of the new cfdmfFTFoam solver is reasonably close to (slightly below) the standard interFoam OF solver. For both solvers, using 2 processors, a nearly linear speedup of about 2 is attained, and the best speedup is achieved for 4 processors, for the current setup. The degradation in efficiency observed at larger processor counts for both solvers can be associated with several factors, including the choice of linear solver for systems of equations, the number of cells per processor, and the hardware used.

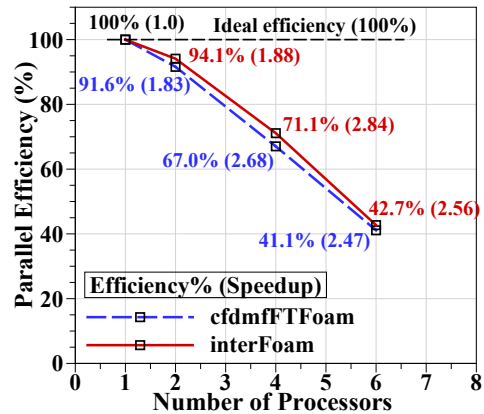

**Figure 10** The droplet in a Poiseuille Stokes flow benchmark: The parallel efficiency and speedup versus the number of processors.

*5.4. Taylor droplet rising in a tube benchmark*

The next example is a validation against the experimental benchmark by Hayashi, et al. [58] for a silicon-oil Taylor droplet rising in a stagnant glycerol-water fluid column within a vertical circular tube of diameter $D = 31\ mm$. Case 19 from this benchmark is chosen, where the carrier and droplet fluid properties are: $\rho_c = 1240\ kg/m^3$, $\rho_d = 970\ kg/m^3$, $\mu_c = 262 \times 10^{-3}\ Pa.s$, $\mu_d = 485 \times 10^{-3}\ Pa.s$, $\sigma = 28 \times 10^{-3}\ N/m$. No data on the volume of the Taylor droplet was reported, however, it was stated that the flow regime as well as the drop nose and rear terminal shapes are independent of the drop length if $d_s/D < 1.6$, where $d_s$ is the equivalent drop diameter, $d_s = (6V_d/\pi)^{1/3}$. For this simulation, the selected initial shape of the drop is shown in Figure 11a, consisting of a cylinder, which is 21.7 $mm$ in diameter and 82.37 $mm$ in height, and two hemispherical end-caps ($d_s/D \sim 1.3$). The computational domain length is chosen to be $L = 350\ mm$ with the no-slip condition at all boundaries. An unstructured grid with hexa-prism cell shapes (see the top inset of Figure 11a) and a resolution of $\Delta_e \sim 1\ mm$ ($D/\Delta_e \sim 31$) along with a CFL number of 0.15 is used. The initial front surface mesh (Figure 11a) has been generated with the OS software SALOME (www.salome-platform.org) in the standard binary STL format and imported to the cfdmfFTFoam solver, using the "stlMeshConvertor" code provided in the present package (in the "tutorials" directory). The terminal Taylor drop shape predicted by the present simulation is compared with the experimental image in Figure 11b. For this simulation, we used an undulation removal frequency of 25 instead of the default value of 50. This is because in our simulation, the initial drop radius was set to be different from the terminal radius, and the drop large acceleration and radial contraction during the initial stage of the rise (see video 1 provided in the supplementary material) leads to a strong recirculating flow in the drop wake and intense upward flow at the center of the front rear area (see Figure 12). This flow makes a small cavity on the front rear. With the default settings, this cavity becomes too steep at about $t = 3.6\ s$ (Figure 12a), which ends in front self-intersection, invalid triangulation, and code crash. By increasing the frequency of the undulation removal, these very sharp edges are avoided while negligibly affecting the other parts of the front (Figure 12b).

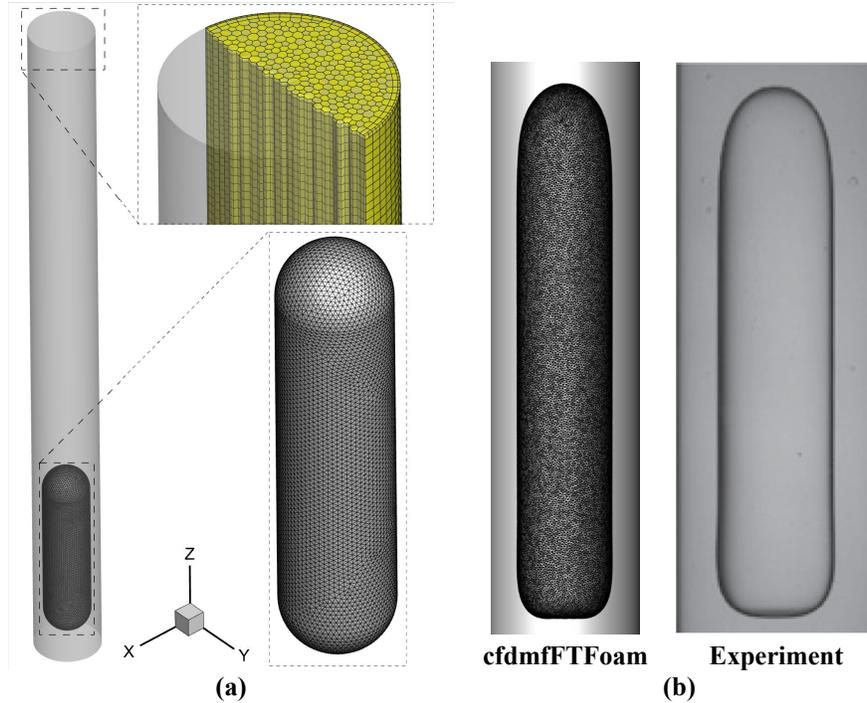

**Figure 11** The Taylor drop rise benchmark: a) The computational domain, a cut of the unstructured hexa-prism Eulerian grid (top inset), and the initial front (bottom inset), b) A comparison of the terminal drop shape predicted by the present FTM solver (at $t = 6\ s$) and the experimental image [58].

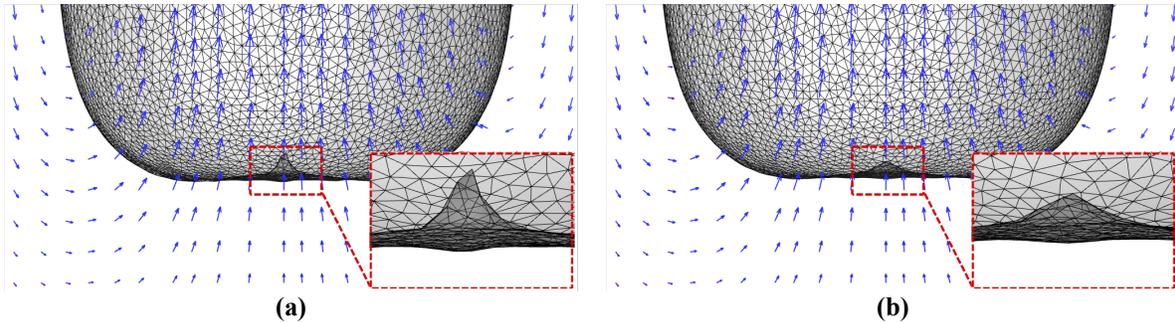

**Figure 12** The rear of the Taylor drop at $t = 3.6\ s$ predicted by the present FTM solver using the undulation removal frequency of a) 50 and b) 25.

## 5.5. Free rising bubble benchmark

For an evaluation of the solver in a challenging problem with a large inertial effect (Reynolds number), an intense interface deformation, and high density and viscosity ratios ($10^3$-$10^4$), the famous 3D rising bubble benchmark by Bhaga and Weber [59] is used, where a spherical bubble of diameter $d = 2.61\ mm$ is released from the rest in a stagnant heavier fluid. The physical properties of the chosen case are: $\rho_c = 1350\ kg/m^3$, $\rho_d = 1.225\ kg/m^3$, $\mu_c = 0.54$ Pa.$s$, $\mu_d = 1.78 \times 10^{-5}$ Pa.$s$, $\sigma = 0.078\ N/m$. The corresponding governing dimensionless parameters are the Eötvös number, $Eo = (\rho_c - \rho_d)gd^2/\sigma = 116$, the Morton number, $Mo = (\rho_c - \rho_d)gd^2\mu_c^4/(\rho_c^2\sigma^3) = 1.31$, and property ratios, $\rho_c/\rho_d = 1102$ and $\mu_c/\mu_d = 3.03 \times 10^4$. For this case, a cuboid computational domain ($8d \times 8d \times 14d$), as shown in Figure 13a, is chosen with a no-slip condition at the bottom wall, a zero-valued pressure and zero-gradient velocity at the top wall, and a zero-gradient velocity and pressure at the side walls. Applying two or three levels of static Eulerian grid refinement in a cuboid of size ($3d \times 3d \times 8d$) results in a grid resolution of $d/\Delta = 20$ or $32$. The simulation is performed with a CFL number of 0.1 and continues till $t\sqrt{g/d} = 6$. Figure 13b shows the rise velocity over time, predicted by the present FTM solver against the other reference numerical solutions available in the literature and the terminal velocity measured in the reference experiment. The terminal bubble shapes are also compared in Figure 13c. While the present FTM accurately predicts the bubble terminal rising velocity (Figure 13b) and captures the overall fully-developed shape and regime of the bubble (Figure 13c), which is classified as an oblate ellipsoidal cap bubble, the level of deformation (the axial shortening and lateral expansion) of the bubble is underpredicted compared to the experiment. This error may be attributed to an inconsistency of the momentum discretization, within the "interFoam" solver of the OF CFD package, used as the basis of the present FTM solver for the pressure-velocity coupling (the PIMPLE loop). Liu, *et al.* [32] pointed out that this kind of inconsistency could lead to a level of error and solver instability at very large density ratios, which is the case for this last benchmark. Therefore, adapting the consistent momentum discretization proposed by Liu, *et al.* [32] to the present FTM solver could enhance its performance at very large property ratios and is a valuable topic for future research.

To study the effects of the kernel radius, regulated by the coefficient $\alpha$ in Eq. (23), simulations with different values of $\alpha$ are carried out, and the results are shown in Figure 14. As can be seen, distributing the surface tension, and density/viscosity gradients over a smaller interfacial thickness, i.e., by a smaller $\alpha$ or a finer grid resolution (larger $d/\Delta$), which is more aligned with the physics of a sharp interface, improves the predictions in terms of both the rise velocity and the axial shortening of the shape profile, leading to better agreement with the experiment. However, we observed that for $\alpha < 2$, the solution diverges without the Tikhonov regularization, described in section 4.5, due to the singularity of the coefficient matrix, which is induced by a too small kernel radius. Therefore, the singularity treatment is essential for a robust usage of RKPM method in the current FTM.

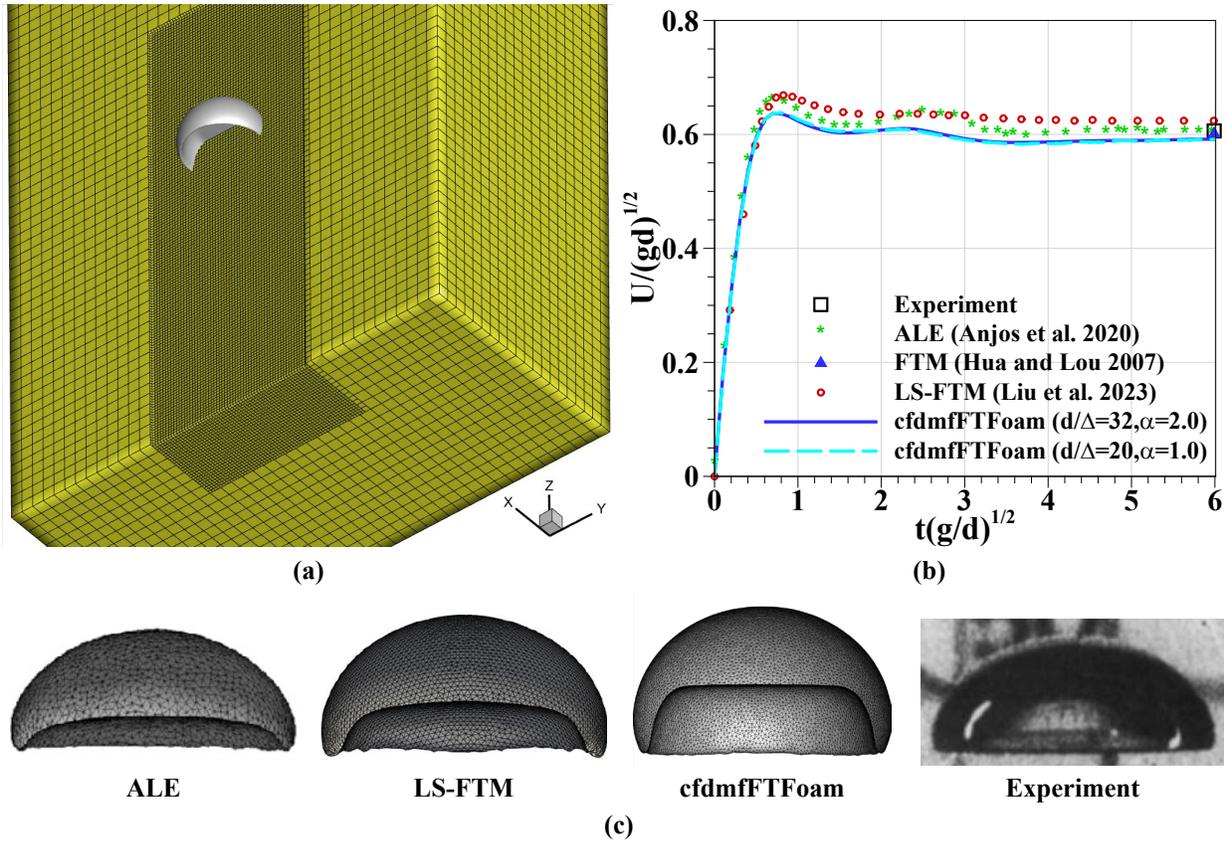

**Figure 13** The free rising bubble benchmark: a) A cut of the Eulerian grid, b) the dimensionless rise velocity versus dimensionless time, and c) the terminal bubble shapes. The comparison of the present "cfdmfFTFoam" FTM simulation (3D), reference numerical simulations, ALE (2D) [60], FTM (2D) [61], LS-FTM (3D) "lentFoam" [32], and experimental measurements [59].

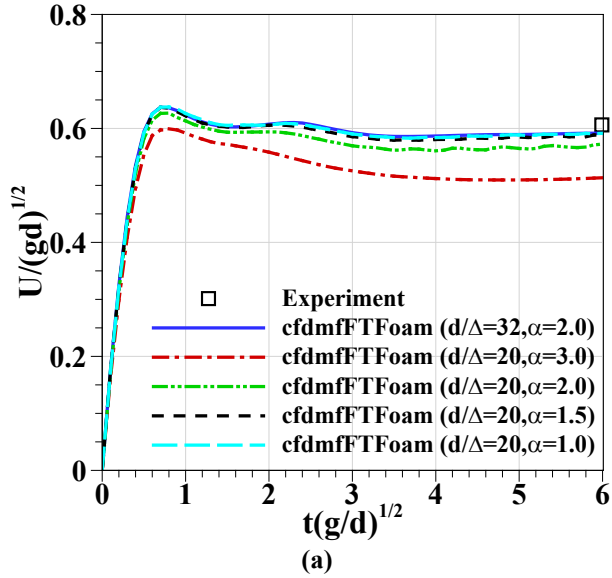

(a)

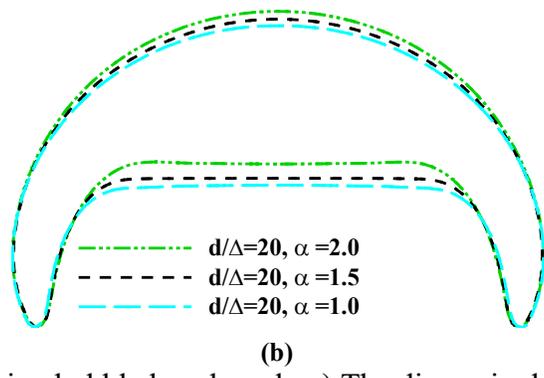

(b)

**Figure 14** The free rising bubble benchmark: a) The dimensionless rise velocity versus dimensionless time, and b) the terminal bubble shape profiles, for different values of the kernel radius coefficient, $\alpha$.

## 6. Conclusion

In the present work, an open-source open-access FTM package, called cfdmfFTFoam, was provided for CFD simulation of flows of incompressible immiscible fluid phases, with the capability of handling problems on general unstructured Eulerian grids and domain-decomposed parallel runs. For this purpose, the Ftc3D FTM code was integrated into the OpenFOAM CFD software, and a variety of additional FTM sub-algorithms, including the front volume correction, front remeshing, surface tension computation, front-to-field communication, and front advection were implemented. The important features of the solver and many of its main underlying algorithms were detailed in the present work. Then, a systematic benchmarking was conducted to validate different sub-algorithms, assess the capability of the new solver in predicting multiphase flow phenomena, and provide examples of case setups and working with different utilities of the package.

The present solver can be used as a basis to ease additional FTM algorithm progress and multiphase flow studies in the future. Possible improvements would be the inclusion of contact-line dynamics, topology change algorithms, interfacial heat and mass transport, and development of new hybrid FTM strategies.

**Declaration of Interests:** The author reports no conflict of interest

**Acknowledgments**

I would like to express my appreciation to Prof. Gretar Tryggvason, Johns Hopkins University, USA, for sharing his codes, giving us opportunities for collaborations on FTM projects during the past two years, and valuable knowledge I gained from him through his seminal publications on FTM. I would like to thank my former CFDMF group students, Mahdi Jafari, Pedram Yousefi, Mohammad Amin Amini, Mohammad Bagher Molaei, and Mehran Sharifi who partly contributed to developing the "cfdmfFTFoam" solver and/or tutorials.

**Data availability**

This paper has been linked with the publicly available GitHub and CPC repositories.

**Appendix A: The package structure and class hierarchies**

The main solver file, i.e., "cfdmfFTFoam.C", is shown in Listing 1. The solver is based on the PIMPLE pressure-velocity coupling, where in each time loop, the FTM sub-algorithms are called in lines 181-190, after a possible Eulerian mesh dynamic refinement (in lines 145-178). The main FTM object `frontTrackingCloud` encapsulates the Lagrangian surface mesh (front) data and FTM procedures. The method `evolve()` in line 184 directs all FTM sub-algorithms shown in Figure 2(right), and in lines 186-188, the mixture density, `rho` ($\rho$), viscosity, `mu` ($\mu$), and surface tension, `sTension` ($F_\sigma$), fields are updated to solve the one-fluid NS equations, Eqs. (1)-(2). The `conservativeSmooth` function in line 189, included through "conservativeSmooth.H" header file, is a conservative filter to smooth surface tension field and increase stability at the cost of decreasing accuracy. This filter has not been used in any of our solutions in section 5 (by setting `nCFilterLoops` to zero). The rest of this code is similar to the standard "interFoam" VOF solver of OF9.

**Listing 1** A summary of important lines of the solver main file. The triple-dot sign "…" shows a part of the code which has not been displayed for the sake of brevity.

```
solvers/…/cfdmfFTFoam/cfdmfFTFoam.C
     ...
     Application
         cfdmfFTFoam
     Description
         Solver for N incompressible, isothermal immiscible fluids using a FTM
         (Front Tracking Method),
         with optional mesh motion and mesh topology changes including adaptive
         re-meshing. (baseCode: interFoam)
     \*---------------------------------------------------------------------------*/
35   #include "fvCFD.H"
     ...
58   int main(int argc, char *argv[])
59   {
         ...
84       while (pimple.run(runTime)) //Time loop
         {
             ...
115          while (pimple.loop()) //PIMPLE loop
116          {
117              if (pimple.firstPimpleIter() || moveMeshOuterCorrectors)
118              {
                     ...
143                  frontTrackingCloud.storeGlobalPositions(); //EA
145                  mesh.update();//Dynamic mesh refinement
                     ...
178              }
180              //EA //Front tracking steps
181              if (pimple.firstPimpleIter())
182              {
183                  Info<< "Evolving " << frontTrackingCloud.name() << endl;
184                  frontTrackingCloud.evolve();
186                  frontTrackingCloud.updateDensityFromFT(rho);
187                  frontTrackingCloud.updateViscosityFromFT(mu);
188                  sTension.primitiveFieldRef() = frontTrackingCloud.sTensionForceFromFT() -
     frontTrackingCloud.pressureJumpAtTheInterfaceFromFT();
189                  if (nCFilterLoops > 0) sTension = conservativeSmooth(sTension, mesh, nCFilterLoops);
190              }
                 ...
208              rhoPhi == fvc::interpolate(rho)*phi; //EA
                 ...
212              #include "UEqn.H"
214              // --- Pressure corrector loop
215              while (pimple.correct()) //PISO loop
216              {
217                  #include "pEqn.H"
218              }
     ...
```

The `frontTrackingCloud` object, instantiated from the class of the same name, is supplied to the solver by the "frontTracking" library. The structure of this library, which is the core of the present FTM, is shown in Figure 15. The UML class diagrams of important classes are also presented in this figure.

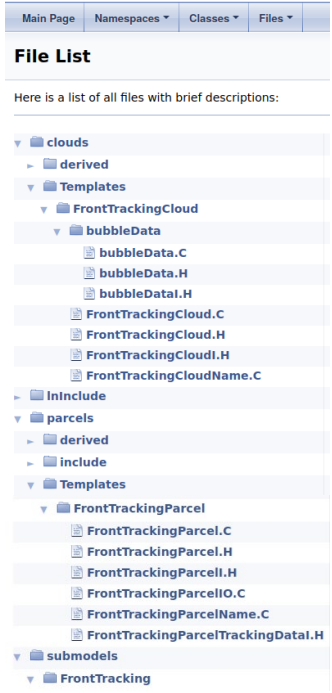
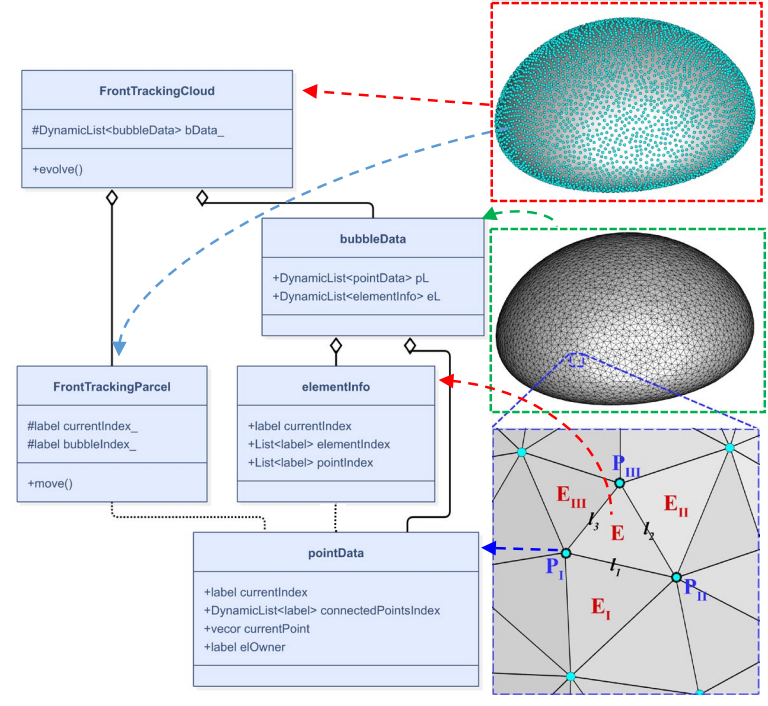
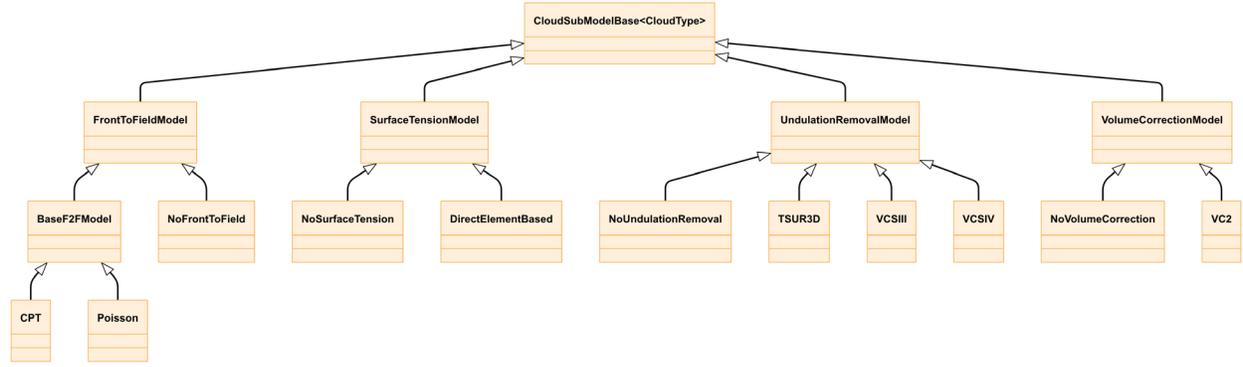

(a)　　　　　　　　　　　　　　　　(b)

(c)

**Figure 15** The "frontTracking" library: a) The file structure, b) important front classes, and c) important sub-model classes.

The most basic class of a surface mesh (or front) is `pointData` which is a vertex (or point) of a front (Figure 15b). Its important member data includes `currentIndex` – the label for identification of the vertex storage location (index) in the list of vertices, `currentPoint` – the vertex position vector, `connectedPointsIndex` – a list of indices of all vertices having a mesh connectivity to the current vertex, and `elOwner` – the index of the front triangular element to which the current vertex belongs. The next class shown in Figure 15b is `elementInfo` which indicates a triangular element of a front. It contains an identifier index in the list of elements, `currentIndex`, a list of indices of the three neighboring elements ($E_I$, $E_{II}$, and $E_{III}$), `elementIndex`, and a list of indices of its three vertices ($P_I$, $P_{II}$, and $P_{III}$), `pointIndex`. In our code, `bubbleData` is a class showing a closed surface entity, like a bubble or drop interface, which contains all information of a Lagrangian surface mesh, such as a list of front vertices (`DynamicList<pointData>`) and elements (`DynamicList<elementInfo>`). A multiphase flow can have multiple interfaces, i.e., a list of `bubbleData`. All these front mesh classes are defined in address "src/…/frontTracking/clouds/derived/frontTrackingCloud/bubbleData/bubbleData.H" (see Figure 15a). To utilize the Lagrangian particle tracking tools of OF, we added new parcel and cloud template classes, called `FrontTrackingParcel` and `FrontTrackingCloud` (Figure 15). A `FrontTrackingParcel` object is used to advect a front vertex location and is associated with a `pointData` object. `FrontTrackingCloud` template class encapsulates the main FTM procedures. The final `frontTrackingCloud` class inherits from `FrontTrackingCloud` and the well-known `MomentumCloud` class of the standard OF to have all their data and functionalities (see Listing 2).

**Listing 2** The definition of the `frontTrackingCloud` class.

```
src/…/frontTracking/clouds/derived/
   frontTrackingCloud/frontTrackingCloud.H
...
43  namespace Foam
44  {
45      typedef ParcelCloud
46      <
47          FrontTrackingCloud
48          <
49              MomentumCloud
50              <
51                  ParcelCloudBase
52                  <
53                      frontTrackingParcel
54                  >
55              >
56          >
57      > frontTrackingCloud;
58  }
...
```

To understand the important methods of the key `frontTrackingCloud` class and the FTM call sequence, an activity diagram of FTM sub-algorithms is reported in Figure 16. The start of the sequence is the call to the `evolve()` method of the `frontTrackingCloud` class, see also line 184 of the main solver in Listing 1. This method contains a call to the `solve()` method of the same class which directs the run eventually to the `motion()` method. This method coordinates the FTM sub-algorithms (shown in Figure 2(right)) by calling `procesingTheFrontMeshes()` – performing the volume correction and remeshing sub-algorithms, `surfaceTensionModel_->calculate()` – performing surface tension computation, `frontToFieldModel_->calculate()` – responsible for front-to-field communication and indicator function construction, and `frontCloudAdvection()` – advecting the front vertices. These models are selected by the OF run-time selection mechanism, from the options shown in Figure 15c. The details of these models are given in sections 4.1-4.6. It is worth mentioning that for an optimized execution time in parallel runs, each processor has a full copy of every front (`bubbleData`), while OF decomposes a cloud (`frontTrackingCloud`) and distributes its parcels between processors. Therefore, special algorithms are required to synchronize cloud parcels and front vertices updates. This is accomplished whenever necessary by `updateParcelsFromFront()` and `mapParcelToFront()` methods (Figure 15). Regarding the `FrontTrackingParcel` template class (Figure 15), we added the `moveDistance()` method for advecting a parcel for a given distance, $\Delta r$, within an Eulerian unstructured grid, in addition to the well-known standard OF `move()` method for advecting parcels for a given time step, $\Delta t$.

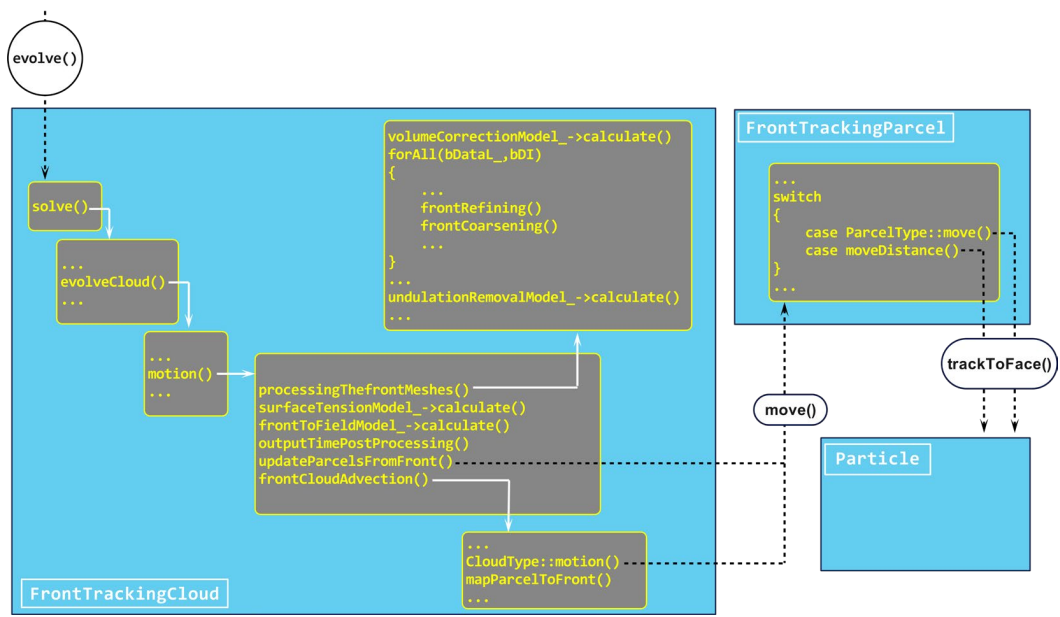

**Figure 16** The activity diagram: The call sequence of important class methods. Each blue box indicates a specific class and each gray box a specific method.

**Appendix B: The volume correction code implementation**

The `calculate()` method of the VC2 algorithm is reported in Listing 3. In line 82 of the listing, a loop is performed over `bubbleData`s to carry out the steps of the algorithm for all fronts. Since the fronts are continuously evolving, their properties are continuously changing. To optimize the code, we used flags to update a front data only when it is required as an input to a current algorithm. Lines 85 and 86 check some flags to see if the inputs of the algorithm, such as the bubble volume (line 1 of Algorithm 1), are up to date, otherwise they are updated. Lines 92 and 93 (line 2 of Algorithm 1) compute the volume error and check a tolerance to apply VC2. Lines 139-170 correspond to lines 3-5 of Algorithm 1 for computing $\boldsymbol{n}_p$, given the positions of all neighboring vertices, stored in the `connectedPoints` data list, of a given vertex, `currentPt`. Lines 172-193 in Listing 3 performs lines 6-9 of Algorithm 1, and lines 195-212 (line 10 of Algorithm 1) solves Eq. (12) using the Newton-Raphson method. Lines 216-221 (line 11-13 of Algorithm 1) correct positions of all vertices. All front sub-algorithms have a line like line 227 which calls a method to let the flags know some front parameters have to be updated (later) due to changes in fronts. For example, in this case, displacements of the front vertices by the VC2 algorithm change the elements' areas, which must be updated before being used as inputs to any algorithm.

**Listing 3** A summary of important lines of the VC2 model `calculate()` method. This method is referenced by `volumeCorrectionModel_->calculate()` in Figure 15.

```
src/.../frontTracking/submodels/FrontTracking/
    VolumeCorrectionModels/VC2/VC2.C
...
67  template<class CloudType>
68  void Foam::VolumeCorrectionModels::VC2<CloudType>::calculate()
69  {
70      Info << "\n--------> Performing the volume correction ... " << endl;
80      DynamicList<bubbleData>& bDataL_ = this->owner().bData();
82      forAll(bDataL_,bDI)
83      {//1
74          //necessary updates
85          if (!bDataL_[bDI].plConnectionFlag) bDataL_[bDI].ptConnectedPoints();
86          if (!bDataL_[bDI].volumeFlag) bDataL_[bDI].updateBubbleVolume();
89          DynamicList<pointData>& pL = bDataL_[bDI].pL;
90          DynamicList<elementInfo>& eL = bDataL_[bDI].eL;
92          scalar Vloss = bDataL_[bDI].volume0 - bDataL_[bDI].volume;
93          if (mag(Vloss)/bDataL_[bDI].volume0 > volumeCorrectionTolerance_)
94          {
95              // normal velocity calculation at points
96              DynamicList<vector> dp;
139             forAll(pL, ptI)
140             {//2
142                 pointData& currentPt = pL[ptI];
143                 point& x = currentPt.currentPoint;
144                 vector dn = vector::zero;
148                 DynamicList<point>& cPoints = currentPt.connectedPoints;
150                 point xn = cPoints[cPoints.size()-1];
151                 point x0 = cPoints[0];
152                 vector normalSum = vector::zero;
153                 normalSum = normalSum + ( (xn - x) ^ (x0 - x) );
154                 for( int cPI = 0; cPI <= cPoints.size()-2; cPI++)
155                 {
156                     point xi = cPoints[cPI];
157                     point xii = cPoints[cPI+1];
158                     normalSum = normalSum + ( (xi - x) ^ (xii - x) );
159                 }
160                 dn = normalSum/(mag(normalSum)+ROOTVSMALL);
169                 dp.append(dn);
170             }//2
172             //Calculating a, b, c
173             scalar a = 0.0;
174             scalar b = 0.0;
175             scalar c = 0.0;
177             forAll(eL, elementI)
178             {//4
179                 elementInfo& elInfo = eL[elementI];
180                 point x1 = pL[elInfo.pointIndex[0]].currentPoint;
181                 point x2 = pL[elInfo.pointIndex[1]].currentPoint;
182                 point x3 = pL[elInfo.pointIndex[2]].currentPoint;
183                 vector n1 = dp[elInfo.pointIndex[0]];
184                 vector n2 = dp[elInfo.pointIndex[1]];
185                 vector n3 = dp[elInfo.pointIndex[2]];
187                 a += (n1 & (n2 ^ n3));
188                 b += (x1 & (n2 ^ n3)) + (x2 & (n3 ^ n1)) + (x3 & (n1 ^ n2));
189                 c += (n1 & (x2 ^ x3)) + (n2 & (x3 ^ x1)) + (n3 & (x1 ^ x2));
190             }//4
191             a /= 6;
192             b /= 6;
193             c /= 6;
195             //Solving for eps in Vloss=a*eps^3+b*eps^2+c*eps
196             scalar eps = 0; //initial guess
197             scalar guess = eps;
198             scalar error = 1.e9;
200             for (label nIter = 0; nIter < VCMaxIter_; ++nIter)
201             {
202                 scalar f = a*pow(eps,3)+b*eps*eps+c*eps-Vloss;
203                 scalar d = 3*a*eps*eps+2*b*eps+c;
205                 eps = eps - f/d;
207                 error = mag(eps - guess)/(mag(eps)+VSMALL);
209                 if (error <= volumeCorrectionTolerance_) break;
211                 guess = eps;
212             }
216             //correct vertices
217             forAll(pL, ptI)
218             {//2
219                 pL[ptI].currentPoint += eps*dp[ptI];
221             }//2
224         }
225     }//1
227     this->owner().setMotionFlags();
228 }
...
```